\documentstyle[psfig]{elsart}

\def\gray{{$\gamma$-ray}}
\def\grays{{$\gamma$-rays}}

\def\mnras{{\it Mon. Not. Roy. Astron. Soc.\ }}
\def\gray{$\gamma$-ray\ }
\def\grays{$\gamma$-rays\ }
\def\apj{{\it Astrophys. J.\ }}
\def\prl{{\it Phys. Rev. Letters\ }}

\def\gray{{$\gamma$-ray}}
\def\grays{{$\gamma$-rays}}

\def\epr{{e-print astro-ph/}}

\begin{document} 

\begin{frontmatter}

\title{Ultrahigh Energy Cosmic Rays: Old Physics or New Physics?}

\author{F.W. Stecker}

\address{Laboratory for High Energy Astrophysics\\
NASA Goddard Space Flight Center, Greenbelt, MD, USA}

\begin{abstract}               

We consider the advantages of and the problems associated with hypotheses
to explain the origin of ultrahigh energy cosmic rays (UHECR: $E > 10$ EeV) 
and the ``trans-GZK'' cosmic rays (TGZK: $E > 100$ EeV) both through 
``old physics'' (acceleration in cosmic sources) and ``new physics'' 
(new particles, topological defects, fat neutrino cross sections, 
Lorentz invariance violation).  

\end{abstract}




\end{frontmatter}

\section{Introduction}

\begin{figure}
\centerline{\psfig{figure=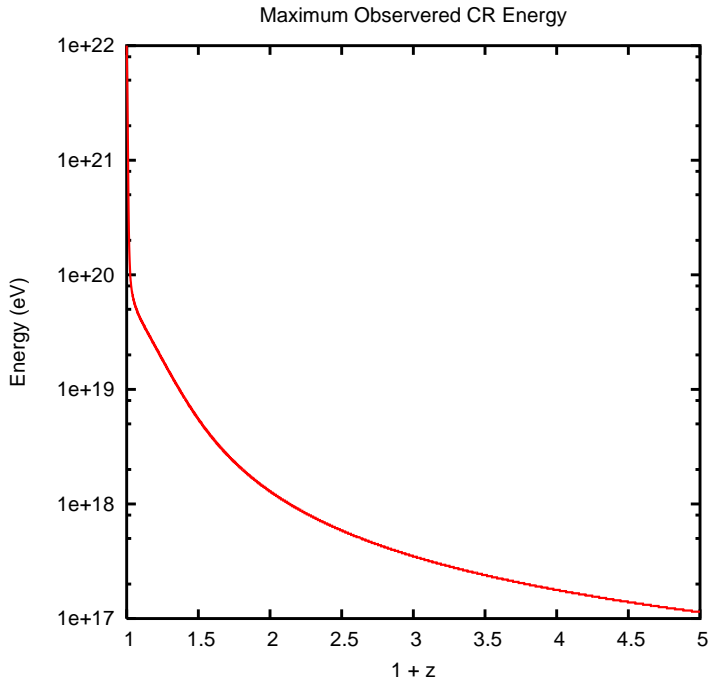,height=14cm}}
\vspace{-1.5cm}
\caption{The GZK cutoff energy, defined as the energy predicted
for a flux decrease of $1/e$ owing to intergalactic photomeson production
interactions, as a function of redshift (Stecker and Scully 2004).}
\label{egzk}
\end{figure}

Owing to their observed isotropy ({\it e.g.,} Stokes, these proceedings), 
and ultrahigh energy, cosmic rays above 10 EeV (1 EeV $\equiv 10^{18}$ eV) 
are believed to be of extragalactic origin. Shortly after the discovery 
of the 3K cosmogenic background radiation (CBR), Greisen (1966) and Zatsepin 
and Kuz'min (1966) predicted that pion-producing interactions of such cosmic 
ray protons with the CBR should
produce a spectral cutoff at  $E \sim$ 50 EeV (the GZK cutoff). The GZK effect 
is not a true cutoff, but a suppression of the ultrahigh energy cosmic ray 
flux owing to an energy dependent propagation time against energy losses by 
such interactions, a time which is only 300 Myr for 100 EeV protons 
(Stecker 1968). At high redshifts, $z$, the target photon density increases by 
$(1+z)^3$ and both the photon and initial cosmic ray energies increase by 
$(1+z)$. A plot of the GZK energy as a function of redshift, 
calculated for the $\Lambda$CDM cosmology, is shown in Figure \ref{egzk} 
(Stecker and Scully 2004). If the source spectrum is hard enough, 
there could also be a relative enhancement just below the ``GZK energy'' 
owing to a ``pileup'' of cosmic rays starting out at 
higher energies and crowding up in energy space at or below the predicted 
cutoff energy. At energies in the 1-10 EeV range, pair production interactions
should take a bite out of the UHECR spectrum.

UHECRs produce giant air showers. Observational studies of these showers
have been done with   
scintillator arrays and with atmospheric flourescence detectors.
The {\it AGASA} scintillator array collaboration claims a significant number 
of events at trans-GZK energies (Teshima, these proceedings). However, 
{\it HiRes} monocular data obtained using the flourescence technique appear to 
be consistent with the GZK effect (Zech, these proceedings). The {\it AGASA} 
data indicate a deviation from the predicted GZK effect,
even if the number density of ultrahigh energy sources is weighted 
like the local galaxy distribution (Blanton, {\it et al.} 2001). 
De Marco {\it et al.}\ (2003) have argued that the discrepency
between the {\it AGASA} and {\it HiRes} results is not statistically 
significant. Better statistics will require data from future ground based
detectors such as {\it Auger} (Suomijarvi, Privatera, and Perrone, 
these proceedings) and space based detectors such as 
{\it OWL} (Stecker, elsewhere in these proceedings) and
{\it EUSO} (D'Ali'Stati, these proceedings). 

The {\it Auger} project now
under construction will use both scintillators and fluorscence detectors so
that  combined results from {\it Auger} 
can help clarify the present {\it prima facie}  
discrepency between the {\it AGASA} and {\it HiRes} results obtained using
these different techiques.. 
(Note that a fluorescence detector such as {\it HiRes}, namely {\it Fly's 
Eye}, reported the highest energy event yet seen, {it viz.}, $E \simeq 300$ 
EeV.) It is apparent that the observational situation is interesting 
enough and the physics implications are important enough to justify both more 
sensitive future detectors and the theoretical investigation
of new physics and astrophysics.
The significance of a non-observation of a GZK effect is profound. Such a 
result either requires a large overdensity of UHECRs within about 100 Mpc 
emitted by unidentified ``local'' sources and trapped by magnetic fields, 
or it requires new physics.

\section{Old Physics: The ``Bottom-Up'' Scenario}

The apparent lack of a GZK cutoff (with the exception of the new {\it HiRes} 
results) has led astrophysicists to hunt for nearby cosmic
``zevatrons'' which can accelerate particles to energies 
$\cal{O}$(1 ZeV $\equiv 10^{21}$eV). 
It is generally assumed 
that the diffusive shock acceleration process is the most likely mechanism for
accelerating particles to high energy in astrophysical sources. 
In this case, the maximum obtainable
energy is given by $E_{max}=keZ(u/c)BL$, where $u \le c$ is the shock speed, 
$eZ$ is the charge of the particle being accelerated, $B$ is the magnetic field
strength, $L$ is the size of the accelerating region and the numerical 
parameter $k = \cal{O}$$(1)$. 
Taking $k = 1$ and $u = c$, one finds $E_{max} = 0.9Z(BL)$, with $E$ in EeV, 
$B$ in $\mu$G and $L$ in kpc. This assumes that 
particles can be accelerated efficiently up until the moment when they can 
no longer be contained by the source, {\it i.e.}, until their gyroradius 
becomes larger than the size of the source.
There are not many cosmic zevatron candidates. Galactic
sources such as white dwarfs, neutron stars, pulsars, and magnetars can be 
ruled out because their galactic distribution would lead to UHECR anisotropies 
and this is not the case. Perhaps 
the most promising potential zevatrons are radio lobes of strong radio 
galaxies (Biermann and Strittmatter 1987). The trick is that such 
sources need to be found close enough to avoid the GZK cutoff. For example,
the nearby radio galaxy M87 may be a source of observed trans-GZK 
cosmic rays (Stecker 1968;  Farrar and Piran 2000). Such 
an explanation would require one to invoke magnetic field
configurations capable of producing a quasi-isotropic distribution of 
trans-GZK protons with energies $>100$ EeV, making this hypothesis 
questionable. However, if the primary particles are nuclei (see Section 2.1), 
it is easier to explain a radio galaxy origin for the two highest energy 
events (Stecker and Salamon 1999). 

\subsubsection{The Dead Quasar Origin Hypothesis}

All large galaxies are suspected to
harbor supermassive black holes in their centers which may have once been
quasars, fed by accretion disks which are now used up. It has been 
suggested that nearby quasar
remnants may be the searched-for zevatrons (Boldt and Ghosh 1999; Boldt
and Lowenstein 2000). This scenario also has potential 
theoretical problems and needs to be explored further. In particular, it
has been shown that black holes which are not accreting plasma cannot
possess a large scale magnetic field with which to accelerate particles
to relativistic energies (Ginzburg and Ozernoi 1964; Krolik 1999).
Observational evidence also indicates that the cores of weakly active
galaxies have low magnetic fields (Falcke 2001 and references therein). 

\subsubsection{The Cosmological Gamma-Ray Burst Origin Hypothesis}

It has also been suggested that cosmological $\gamma$-ray bursts (GRBs)
could be the zevatron sources of the highest energy cosmic rays 
if these objects emitted the same amount of energy in
ultrahigh energy ($\sim 10^{14}$ MeV) cosmic rays as in $\sim$ MeV photons
(Waxman 1995; Vietri 1995). 
However, 26 of the 27 bursts with identified host galaxies as of 2003 are 
at moderate to high redshifts ($z > 0.36$),
with GRB00013 having a redshift of 4.50; they are not nearby sources.

The host galaxies of GRBs are sites of very active star formation (Christensen,
{\it et al.}\ 2004). The  bursts occur within star forming regions. 
The GRB redshift distribution follows the strong redshift evolution of the 
cosmic star formation rate (Schmidt 2001: Stern {\it et al.} 2002) with a much 
lower burst rate at the 
low redshifts from which the TGZK events must come. GRBs are thought to be
supernovae caused by the core collapse of massive stars (Cherepashchuk and 
Postnov 2001) and the core collapse supernova rate rate at $z = 0.26$ has 
been found to be a factor of $\sim 3$ higher than the estimate for $z = 0$ 
(Cappellaro, {\it et al.} 2004). There is also some 
evidence for luminosity evolution; GRBs may have been brighter at higher 
redshifts (Amati 2004). 
Schmidt (2001) concludes that the local ($z = 0$) total energy release rate 
by all GRBs in the \gray\ range is $\cal{O}$$(10^{28})$ W Mpc$^{-3}$.
whereas the required energy input rate in UHECRs above 10
EeV  is  $\cal{O}$$(10^{31})$ W Mpc$^{-3}$.
GRBs fail by at least an order of magnitude to account
for TGZK ($>100$ EeV) events 
(Stecker 2000) and they fail by
at least two orders of magnitude to account for the UHECR ($>10$ EeV) events 
(Scully and Stecker 2002).

Norris (2002) has given an analysis of the luminosities and
space densities of nearby low luminosity long-lag GRB sources which
are identified with Type I supernovae. For these sources, he finds a rate
per unit volume of $7.8 \times 10^{-7}$ Mpc$^{-3}$yr$^{-1}$ and an average
(isotropic) energy release per burst of 1.3 $\times 10^{49}$ erg over the
energy range from 10 to 1000 keV. The energy release per unit volume is then
$\sim 3 \times 10^{28}$ W Mpc$^{-3}$, more than two orders of
magnitude below the rate needed to account for the TGZK events.
Even these numbers are most likely too optimistic, since they 
are based on the questionable assumption of the same amount of GRB
energy being put into ultrahigh energy cosmic rays as in $\sim$ MeV photons.

\subsection{The Heavy Nuclei Origin Scenario}

\begin{figure}
\centerline{\psfig{figure=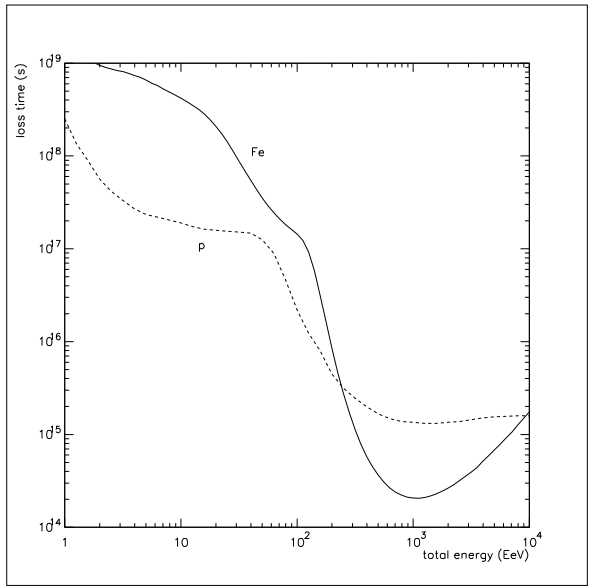,height=10cm}}
\vspace{-1.5cm}
\caption{Mean energy loss times for protons and nuclei originating as iron  
(Stecker and Salamon 1999).}
\label{heavy}
\end{figure}

A more conservative hypothesis for explaining the trans-GZK events is that they
were produced by heavy nuclei. The conditions under which they were 
accelerated in astrophysical sources would have to preclude dissociation. 
Stecker and Salamon (1999) have shown that the energy loss time for nuclei 
starting out as 
Fe is longer than that for protons for energies up to a total energy of 
$\sim$300 EeV (see Fig. \ref{heavy}). Stanev\ {\it et al.} (1995) and Biermann (1998) 
have examined the arrival directions of the highest energy events.
They point out that the $\sim 200$ EeV event is within 10$^\circ$ of the 
direction of the strong radio galaxy NGC 315. This galaxy lies at a distance 
of only $\sim$ 60 Mpc from us. For that distance, the results of Stecker and 
Salamon (1999) indicate that heavy nuclei would
have a cutoff energy of $\sim$ 130 EeV, which may be within the uncertainty in
the energy determination for this event. The $\sim$300 EeV event is within
12$^\circ$ of the direction of the strong radio galaxy 3C134. The distance 
to 3C134 is unfortunately unknown because its location behind a dense 
molecular cloud in our own galaxy obscures the spectral lines required for a 
measurement of its redshift. 
A clue that we may be seeing heavier nuclei above the
proton-GZK cutoff comes from a recent analysis of inclined air showers
above 10 EeV energy (Ave, {\it et al.} 2000). These results favor
proton primaries below the p-GZK cutoff energy but they {\it appear to favor a 
heavier composition above the p-GZK cutoff energy}. We note that continuation 
of the UHECR spectrum to energies significantly above 
300 EeV would rule out heavy nuclei (Stecker and Salamon 1999).

\section{New Physics}

The existence of TGZK events, as well as the difficulty in finding reasonable
candidates for zevatrons, has stimulated theorists to look for possible 
solutions involving new physics. Some of these involve (A) Top-down scenarios 
involving such concepts as grand unification and early universe physics, 
(B) a large increase in the neutrino-nucleon cross section at ultrahigh 
energies, (C) new particles, and (D) a small violation of Lorentz invariance.
It should also be noted that, even if the GZK effect is seen, top-down 
scenarios predict the reemergence of a new component at even higher energies 
(Bhattacharjee and Sigl 2000).

\subsection{Top-Down Scenarios: ``Fraggers''}

A way to avoid the problems with finding plausible astrophysical zevatrons is
to start at the top, {\it i.e.}, the energy scale associated with grand
unification theories (GUTs), supersymmetric grand unification (SUSY-GUTs) and
superstring theory unification. In the very early stages of the big bang,
the universe is believed to have reached temperatures appropriate to
unification theories. 
Very heavy ``topological defects" may be produced as a consequence of
the GUT phase transition when the strong and electroweak forces became
separated. Topological defects in the vacuum of
space are caused by misalignments of the heavy Higgs fields in regions which
were causally disconnected in the early history of the universe. These are 
localized regions where extremely high densities of mass-energy consisting
of early universe Higgs fields are trapped. 
Such defects go by designations such as cosmic strings, monopoles, walls,
necklaces (strings bounded by monopoles), and textures, depending on their 
geometrical and topological properties.   
Superheavy particles or topological structures arising at the GUT energy scale
$M \ge 10^{5}$ EeV can decay or annihilate to produce ``$X$-particles'' (GUT 
scale Higgs particles, superheavy fermions, or leptoquark bosons of mass M.) 
These $X$-particles will 
decay to produce QCD fragmentation jets at ultrahigh energies. I will 
call such sources ``fraggers''. Fraggers produce mainly pions, 
with a 3 to 10 per cent admixture of baryons, so that generally one can 
expect them to produce more high energy $\gamma$-rays and $\nu$'s than 
protons. The number of GUT and SUSY-GUT top-down models
is quite large (Bhattacharjee and Sigl 2000).

\subsubsection{``Z-bursts''}

In principle, ultra-ultrahigh energy $\cal{O}$(10 ZeV) neutrinos
can produce ultrahigh energy $Z$-boson fraggers by interactions with  
1.9K thermal CBR neutrinos (Weiler 1982) resulting in ``$Z$-burst'' 
fragmentation jets. This will occur at the resonance energy $E_{res} =
4[m_{\nu}({\rm eV})]^{-1}$ ZeV. A typical $Z$ boson will decay to produce
$\sim$2 nucleons, $\sim$20 $\gamma$-rays and $\sim$ 50 neutrinos, 2/3 of 
which are $\nu_{\mu}$'s.
If the nucleons produced from $Z$-bursts originate within a few tens of
Mpc of the Earth they can reach us, even though the original $\sim$ 10 ZeV 
neutrinos could have come from a much further distance. 
It has been suggested that this effect can be amplified if our galaxy has
a halo of neutrinos with a mass of tens of eV (Fargion, {\it et al.}\ 1999;
Weiler 1999). However, a neutrino mass large enough 
to be confined to a galaxy size neutrino halo (Tremaine and Gunn 1979) or
even a galaxy cluster size halo (Shafi and Stecker 1984) is now clearly ruled 
out by the results of the Wilkonson Microwave Anisotropy Probe ({\it WMAP}). 
These results, combined with other cosmic microwave background data and
data from the 2dF galaxy redshift survey, together with the very small 
neutrino flavor mass differences 
implied by the atmospheric and solar neutrino oscillation results,
indicate that even the heaviest neutrino 
would have a mass in the sub-eV range, {\it i.e.},
0.03 eV $\le m_{3} \le$ 0.24 eV (Bhattacharyya {\it et al.}\ 2003; Allen {\it et al.}\ 
2003). The tritium decay spectral endpoint
limits on the mass of the $\nu_{e}$  are also consistent with this conclusion. 
Thus, neutrino masses are too small for halo or galaxy cluster confinement.

The severe problem with the $Z$-burst explanation
for the TGZK events is that one needs to produce large fluxes of 
neutrinos with energies in excess of 10 ZeV. If these are secondaries 
from pion production, the primary protons which produce 
them must have energies of hundreds of ZeV! We know of no  
source capable of accelerating particles such energies.
A more likely process to produce 10 ZeV neutrinos would be {\it via} 
top-down fraggers. The flux of such neutrinos is
constrained because the related energy release into electromagnetic 
cascades which produce GeV range \grays\ is limited by satellite 
observations (Bhattacharjee and Sigl 2000). 
This constraint, together with the low probability for 
$Z$-burst production rule out this scenario for explaining the TGZKs.

\subsubsection{Superheavy Dark Matter Particles}

 The inflation of the early universe in the accepted big-bang model
is postulated to be caused by a putative vacuum field
called the inflaton field. During inflation, the universe is cold but, 
when inflation is over, coherent oscillations of the inflaton field reheat 
it to a high temperature. While the inflaton field is oscillating, non-thermal 
production of very heavy particles may take place.  
These heavy particles may survive to the present as dark 
matter. They are also fraggers. Their decays or annihilation will produce 
ultrahigh energy particles and photons {\it via} fragmentation.
It has been suggested that such particles may be the source of ultrahigh
energy cosmic rays (Berezinsky {\it et al.}\ 1997; Kuz'min and Rubakov 1998; 
Blasi {\it et al.} 2002; Sarkar and Toldr\`{a} 2002; Barbot and Drees 2002). 
A comparison of recent experimental constraints from dark matter nuclear 
recoil searches with predicted rates gives a lower limit on the mass of
superheavy dark matter particles of $10^6$ EeV, unless they interact 
weakly with normal matter (Albuquerque amd Baudis 2003).
The annihilation or decay of such particles in a
dark matter halo of our galaxy would produce ultrahigh energy nucleons
which would not be attenuated at TGZK energies owing to their proximity.
The resulting air shower distribution would then be anisotropic.
This would be an even larger effect in the case of annihilation than decay, 
since the flux would then scale as the square of the particle density
density rather than linearly. Since the galactic center is viewed 
from the southern hemisphere, the location of the {\it AUGER} detector will 
make it ideal for testing this hypothesis. 

\subsubsection{Halo Fraggers and the Missing Photon Problem}

Halo fragger models such as $Z$-burst and superheavy halo dark matter 
decay or annihilation will produce
more ultrahigh energy photons than protons. These ultrahigh energy photons 
can reach the Earth from anywhere in a dark matter galactic halo because 
there is a ``mini-window'' for the transmission of ultrahigh 
energy cosmic rays between $\sim 0.1$ and $\sim 10^{6}$ EeV (Stecker
2003). Such photon-induced giant air showers have an evolution profile which 
is significantly different from nucleon-induced showers because of the
Landau-Pomeranchuk-Migdal effect and also cascading in the Earth's 
magnetic field (Cillis\ {\it et al.} 1999). Taking this into account, Shinozaki, {\it
et al.} (2002) have used the AGASA data to place upper limits on 
the primary photon composition of their UHECR events. They find an initial photon 
fraction upper limit of 28\% for events above 10 EeV and 67\% for events above 30 EeV 
at a 95\% confidence level with no indication of photonic showers above 100 
EeV. A recent reanalysis of the ultrahigh energy events observed at Haverah 
Park by Ave, {\it et al.} (2000) 
indicates that less than half of the events (at 95\% confidence level)
observed above 10 and 40 EeV are $\gamma$-ray initiated.
An analysis of the highest energy Fly's Eye event ($E = 300$ EeV) 
shows it not to be of photonic origin (Halzen and Hooper 2002).

In order to solve the missing photon problem for halo fraggers, 
Chisholm and Kolb (2004) have suggested
that a small violation of Lorentz invariance could allow ultrahigh
energy photons to decay into electron-positron pairs (Coleman and Glashow 1999), 
thus eliminating the photon component of the fragger-produced flux.  
The amount of Lorentz invariance
required is within the observational limits obtained by Stecker and Glashow
(2001). However, the scenario suggested by Chisholm and Kolb, 
implies that neutrons would be the primary ultrahigh
particles producing the giant air showers, again producing a halo anisotropy
for which there is no present indication (Shinozaki, {\it et al.} 2002;
Kachelrie\ss\ and Semikoz 2003).

\subsection{Increasing the Neutrino-Nucleon Cross Section at Ultrahigh 
Energies}

Various processes have been invoked to produce observable fluxes of high 
energy neutrinos (see, {\it e.g.}, Stecker 2003). 
Since neutrinos can travel through the universe without interacting with the
2.7K CBR, it has been suggested that if the neutrino-nucleon cross section 
were to increase to hadronic values at ultrahigh energies, they could produce 
the giant air showers and account for the observations of showers above the
proton-GZK cutoff (see, {\it e.g.}, Ringwald, these proceedings).

Several suggestions have been made for processes that can
enhance the neutrino-nucleon cross section
at ultrahigh energies. These suggestions include composite models of neutrinos
(Domokos and Nussinov 1987),
scalar leptoquark  resonance channels (Robinett 1988) and the exchange of dual 
gluons (Bordes, {\it et al.} 1998). Burdman, {\it et al.}\ (1998) have 
ruled out a fairly general class of these types of models
by considering accelerator data and unitarity bounds.
More recently, the prospect of enhanced neutrino cross sections has been 
explored in the context of extra dimension models invoked by some theorists 
as a possible way for accounting for the extraordinary weakness of the 
gravitational force (Arkani-Hamed, {\it et al.}\ 1999; Randall and 
Sundrum 1999). 
These models allow the virtual exchange of gravitons propagating 
in the bulk ({\it i.e.} in the space of full extra dimensions) 
while restricting the propagation of other particles to the familiar four 
dimensional space-time manifold. It has been suggested that in such models,
$\sigma$($\nu$N) $\simeq [E_{\nu}/(100\rm EeV)]$ mb (Nussinov and Schrock
1999; Domokos and Kovesi-Domokos 1999; Jain, {\it et al.} 2000). 
Other scenarios involve the neutrino-initiated
atmopheric production of black holes (Anchordoqui\ {\it et al.}\ 2002; Feng and
Shapere 2002) and higher dimensional extended objects, p-dimensional branes called 
``p-branes'' (Ahn, {\it et al.}\ 2002; Anchordoqui, {\it et al.}\ 2002).
Such interactions, in principle, can increase the neutrino total atmospheric 
interaction cross section by orders of magnitude above the standard model
value. However, sub-mm gravity experiments and astrophysical constraints rule
out total $\nu$N cross sections as large as 100 mb as would be
needed to fit the trans-GZK energy air shower profile data. Nonetheless, extra
dimension models may produce significant increases in this
cross section, resulting in moderately penetrating air showers. Such 
showers should also be present at lower energies
(Anchordoqui\ {\it et al.} 2001; Tyler, {\it et al.}\ 2001). 
No such showers have been observed, putting an indirect 
constraint on extra dimension TeV-scale gravity models.

\subsection{New Particles}

The suggestion has also been made that undiscovered neutral hadrons containing
a light gluino could be producing the trans-GZK events (Farrar 1996;
Cheung, {\it et al.}\ 1998; Berezinsky, {\it et al.}\ 2002). 
While the invocation of such particles is an intriguing idea, it seems 
unlikely that such particles of a few
proton masses would be produced in copious enough quantities in astrophysical
objects without being detected in terrestrial accelerators. There
are also strong accelerator constraints on light gluino production 
(Alavi-Harati, {\it et al.} 1999).
One should note that while it is true that the GZK threshold for such 
particles would be higher than that for protons, 
such is also the case for the more prosaic heavy nuclei previously
discussed. In addition, such neutral particles cannot be accelerated 
directly, but must be produced as secondary particles, making the energetics
reqirements more difficult.

\subsection{Violating Lorentz Invariance}  

With the idea of spontaneous symmetry breaking in particle physics came the
suggestion that Lorentz invariance (LI) might be weakly broken at high energies
(Sato and Tati 1972). Although no true quantum theory of gravity exists, it 
was suggested that LI might be violated in such a theory
with astrophysical consequences (Amelino-Camilia {\it et al.} 1998; Galante,
these proceedings). A simple formulation
for breaking LI by a small first order perturbation in the electromagnetic 
Lagrangian which leads to a renormalizable treatment has been given by
Coleman and Glashow (1999). Using this formalism, these authors 
point out that different particles can have different maximum attainable 
velocities (MAVs) which can be different from $c$. If we denote the MAV
of a particle of type $i$ by $c_{i}$ and the difference $c_{i} - c_{j}
\equiv \delta_{ij}$ then Coleman and Glashow have shown that for 
interactions of protons with CBR photons of energy $\epsilon$ and 
temperature $T_{CBR} = 2.73 K$, pion production is kinematically forbidden and 
thus {\it photomeson interactions are turned off} if

$$\delta_{p\pi} > 5 \times 10^{-24}(\epsilon/T_{CBR})^2.$$

The corrsponding condition for suppression of electron-positron pair production
interactions is given by

$$\delta_{ep} > 5 \times 10^{-19}(\epsilon/T_{CBR})^2.$$

\begin{figure}
\centerline{\psfig{figure=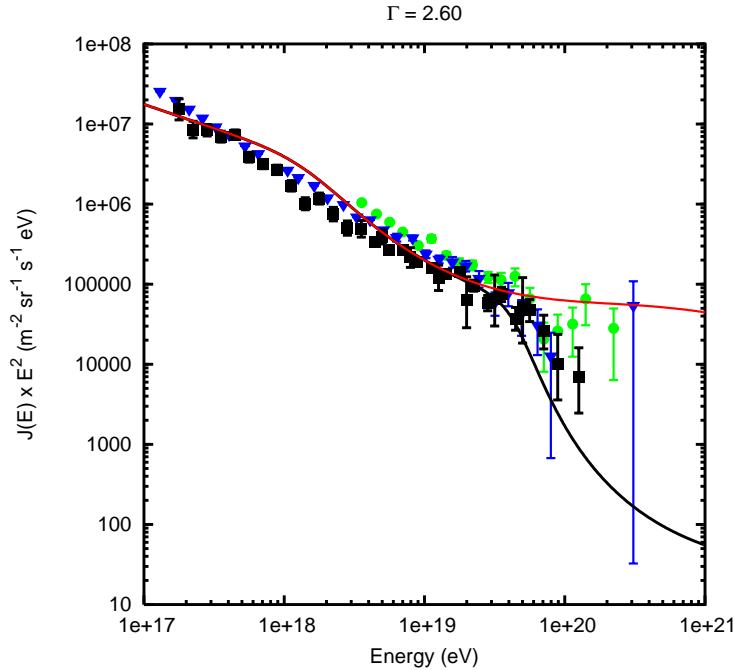,height=14cm}}
\vspace{-1.5cm}
\caption{Predicted spectra for an $E^{-2.6}$ source spectrum with 
source evolution
(see text) shown with pair-production losses included and photomeson 
losses both included (black curve) and turned off (lighter (red) curve) 
(Stecker and Scully 2004). The curves
are shown with ultrahigh energy cosmic ray spectral data from
{\it Fly's Eye} (triangles), {\it AGASA} (circles), 
and {\it HiRes} monocular data (squares). They are normalized to the data
at 3 EeV (see text).} 
\label{f2}
\end{figure}

\begin{figure}
\centerline{\psfig{figure=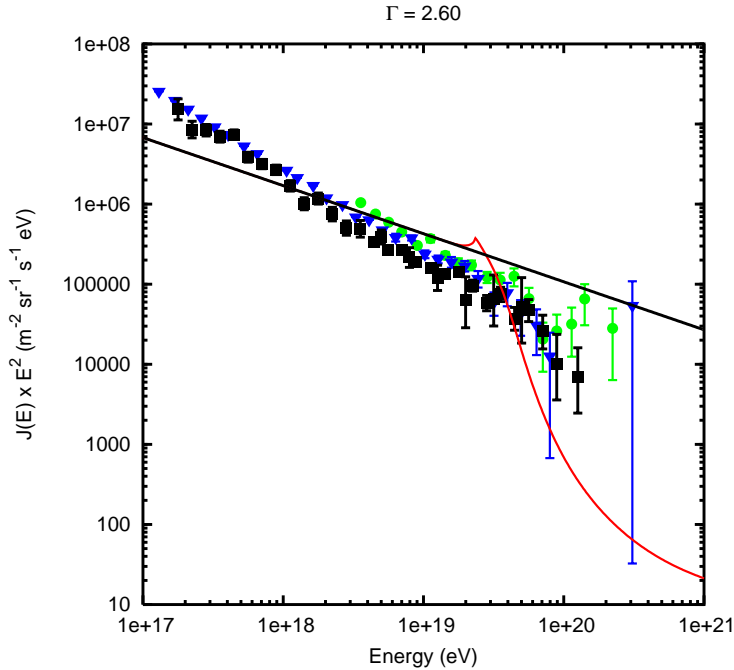,height=14cm}}
\vspace{-1.5cm}
\caption{Predicted spectra for an $E^{-2.6}$ source spectrum with source
evolution
(see text) shown with pair-production losses turned off and with
photomeson production losses included (lighter (red) curve) and turned off
(black straight line) 
(Stecker and Scully 2004). The curves
are shown with ultrahigh energy cosmic ray spectral data from
{\it Fly's Eye} (triangles), {\it AGASA} (circles), and {\it HiRes} monocular 
data (squares). They are normalized to the data at 3 EeV (see text).} 
\label{f3}
\end{figure}

Thus, given even a very small amount of LI violation, both photomeson and 
pair-production interactions of UHECR with the CBR can be turned off. 
Such a violation of Lorentz invariance might be produced by
Planck scale effects (Aloisio, {\it et al.}\ 2000; Alfaro and Palma 2002, 2003). 
The amount of LI violation required is small compared to the constraint 
obtained by Stecker and Glashow (2001) from the non-suppression of 
intergalactic absorption of multi-TeV $\gamma$-rays, {\it viz.}, 
$\delta_{e\gamma} < 1.3 \times 10^{-15}$.
The most stringent constraints to date on LI violation in QED interactions 
are given by Jacobson, {\it et al.}\ (2004).

Figures \ref{f2} and \ref{f3}, show the predicted 
spectra obtained assuming an $E^{-2.6}$
source spectrum and a source luminosity evolution $\propto (1+z)^{3.6}$ 
for $0 < z < 2$ with no further evolution out to $z_{max} = 5$, following the
star formation rate. The resulting spectra, normalized at an energy of 3 EeV
above which energy the extragalactic component is assumed to be dominant
(see, {\it e.g.}, Stecker 2003), are calculated for ``on'' and ``off'' 
energy losses for both photomeson 
production and pair production for protons,

The cosmic ray spectral data from the 
{\it Fly's Eye, AGASA,} 
and {\it HiRes} detectors are also shown.\footnote{Other UHECR data are
given elsewhere in these proceedings and in Nagano and Watson (2000).}
From Figs. \ref{f2} and \ref{f3}, it can be seen that, in principle, 
a very small amount of LI violation can eliminate the GZK ``cutoff''. When 
pair-production is turned off, the $\sim$ 10 EeV ``bite'' in the predicted 
spectrum  is eliminated. Of course, when both interactions are turned off, 
all of the features in the predicted spectrum disappear and only a power-law 
remains. Contrary to the discussion of Alfaro and Palma (2003), 
as can be seen from the curves in Figs. \ref{f2} and \ref{f3}, the present 
data cannot be used to put constriants on LI violation in pair-production 
interactions; 
these data appear to be consistent with either the presence or absence 
of such interactions. In the case of bottom-up models, there remains the
problem of accelerating protons to TGZK energies in the sources (see Section
2) which, of course, is not a problem for top-down models (see Section 3
and Aloisio, these proceedings).

\section{Distinguishing Old {\it vs.} New Physics Scenarios}

Future data which will be obtained with new detector arrays and satellites
will give us more clues relating to the origin of the
trans-GZK events by distinguishing between the various hypotheses that have
been proposed.

An ``old physics'' zevatron origin will produce air-showers primarily
from primaries which are protons or heavy nuclei, with a much smaller 
number of neutrino-induced showers, the neutrinos being secondaries from 
CBR photomeson interactions. Zevatron events should cluster near the direction 
of the sources.

A ``new physics'' fragger origin mechanism will not produce any  
nuclei heavier than protons and will produce more ultrahigh energy neutrinos
than protons. Thus, it will be important to look for the $\nu$-induced air
showers which are expected to originate much more deeply in the atmosphere than
proton-induced air showers and are therefore expected to be mostly
horizontal showers. Such models also produce more photons than protons.
Photons produced in the galactic halo, {\it e.g.}, from 
the decay or annihilation of superheavy dark matter, can reach us 
and will have an anisotropic distribution on the sky.
New physics top-down mechanisms may produce harder spectra than are
expected from cosmic zevatrons. If differential
cosmic ray spectra are parametrized to be of the form $F \propto E^{-\Gamma}$,
then for top-down models $\Gamma < 2$, whereas for bottom-up models
$\Gamma \ge 2$.
If Lorentz invariance violation is the explanation for the missing GZK effect,
one can also look for the absence of a pair-production
10 EeV ``bite'' in the spectrum, but this may be more difficult to detect.

\end{document}